\documentclass[apj]{emulateapj}

\newcommand{\ignore}[1]{}
\usepackage{apjfonts} \usepackage{amsmath,esint} \usepackage{lipsum}
\usepackage{ccaption}
\usepackage{mathtools}
\shorttitle{Characterizing `Oumuamua}
\shortauthors{Seligman \& Laughlin}

\begin{document} \title{The Feasibility and Benefits of In Situ Exploration of `Oumuamua-like objects}

\author{Darryl Seligman\altaffilmark{1}, Gregory Laughlin\altaffilmark{1}, }

\altaffiltext{1}{Department of Astronomy, Yale University, darryl.seligman@yale.edu} 

\begin{abstract}
A rapid accumulation of observations and interpretation have followed in the wake of 1I `Oumuamua's passage through the inner Solar System. We briefly outline the consequences that this first detection of an interstellar asteroid implies for the planet-forming process, and we assess the near-term prospects for detecting and observing (both remotely and \textit{in situ}) future Solar System visitors of this type. Drawing on detailed heat-transfer calculations that take both `Oumuamua's unusual shape and its chaotic tumbling into account, we affirm that the lack of a detectable coma in deep images of the object very likely arises from the presence of a radiation-modified coating of high molecular weight material (rather than a refractory bulk composition). Assuming that `Oumuamua is a typical representative of a larger population with a kinematic distribution similar to Population I stars in the local galactic neighborhood, we calculate expected arrival rates, impact parameters and velocities of similar objects and assess their prospects for detection using operational and forthcoming facilities. Using `Oumuamua as a proof-of-concept, we assess the prospects for missions that intercept  interstellar objects (ISOs) using conventional chemical propulsion. Using a ``launch on detection'' paradigm, we estimate wait times of order 10 years between favorable mission opportunities with the detection capabilities of the Large-Scale Synoptic Survey Telescope (LSST), a figure that will be refined as the population of interstellar asteroids becomes observationally better constrained.
\end{abstract}


\maketitle

\section{Introduction}\label{introduction}

The interstellar asteroid 1I/2017 U1 (`Oumuamua), presented astronomers with their first glimpse of an object entering the inner Solar System with a hyperbolic trajectory. The detection of this remarkable body was made on October 19th, 2017 by the Panoramic Survey Telescope and Rapid Response System (Pan-STARRS) project \citep{Chambers2016}, and its extrasolar origin was rapidly confirmed, leading to an October 25th, 2017 discovery announcement through the Minor Planet Center \citep{Mpec2017a}. Its securely determined perihelion distance, $p=0.255\,{\rm AU}$, and eccentricity, $e=1.2$, imply a pre-encounter velocity, $v_\infty=26\,{\rm{km/s}}$. No known or proposed outer Solar System objects are remotely capable of imparting such an impulse; `Oumuamua is a singular outsider.

In the weeks following `Oumuamua's discovery, observations by many groups provided significant additional detail concerning its properties. The near-daily flux of articles was large enough that a review of the current state of knowledge is now useful.

Moderate-resolution spectra of `Oumuamua were obtained by \citet{Masiero2017} and \citet{Ye2017}  immediately after the Minor Plant Center's circulation of the discovery announcement. These measurements showed no sign of narrow emission or absorption features, and pointed to a spectral gradient, $S^{\prime}$ that is skewed toward the red. \citet{Ye2017} measured a gradient $S^{\prime}=0.1\pm0.06\, ({\rm100\, nm})^{-1}$ at 650 nm, based on a 400-900 nm band pass. This value for $S^{\prime}$ is consistent with that displayed by cometary nuclei, Trojan asteroids and inactive Centaurs, but is less red than active Centaurs and all classes of Kuiper Belt Objects, which typically exhibit ultra-red gradients $S^{\prime}=0.23\pm0.02\, ({\rm100\, nm})^{-1}$ in the optical region \citep{Jewitt2015}. \citet{Bannister2017} obtained $g^{\prime}$, $r^{\prime}$, and $J$ photometry, which extended `Oumuamua's color assessment beyond 1$\,\mu$m and slightly into the near-infrared. Their results indicate a spectral slope $S^{\prime}=0.22\pm0.15\, ({\rm100\, nm})^{-1}$, consistent within the errors to the results obtained in the visible region of the spectrum. Additionally, \citet{Ye2017} searched radar meteor data taken during the relevant arrival time and found that no incoming meteors appeared to have been associated with `Oumuamua (which passed only 0.161 AU from Earth at close approach).

The arrival direction of `Oumuamua's incoming trajectory was located close to the solar apex (the direction of the Sun's peculiar motion with respect to the Local Standard of Rest), providing strong circumstantial evidence that it is a member of a large population of planetesimals ejected from planetary systems throughout the Milky Way's disk during their formation phases \citep{Laughlin2017, Trilling2017}. Its kinematics suggest that it belongs to a population with kinematics similar to the Population I stars of the solar neighborhood \citep{Mamajek2017}. Indeed, \citet{Wright2017} reiterated that known (or unknown) planets in our own Solar System have no chance of scattering `Oumuamua from an Oort Cloud origin into its hyperbolic trajectory. 

Several authors have discussed the possibility that specific source stars of origin for `Oumuamua might be identifiable. \citet{Gaidos2017a} draw on velocity similarities to suggest that `Oumuamua was formed $\sim\,$45 Myr ago in a protoplanetary disk of one of the 50-85 pc-distant stars of the Carina/Columba association.
\citet{Zuluaga2017} presented a general method to assign probabilities of specific nearby stars being parent systems for interstellar asteroids, and presented their most likely source candidate for `Oumuamua as the double system HD~200325. \citet{Dybczynski2017} carried out a detailed dynamical assessment of more than two hundred thousand well-characterized nearby stars, with the aim of identifying potential source systems, and found no particularly compelling candidates. Intriguingly, Dybczy{\'n}ski \& Kr{\'o}likowska do note that roughly 820,000 years ago, `Oumuamua encountered the nearby planet-bearing star Gliese 876 \citep{Rivera2010} with a $v_\infty=5\,{\rm{km/s}}$ relative velocity, and an impact parameter, $b\sim2.2$ pc. \citet{Dybczynski2017} conclude that this encounter likely occurred at too large a distance to admit a physical connection when the various uncertainties are taken into account. \citet{Zhang2017} emphasized the general difficulties associated with backtracing interstellar objects to their source systems, noting in particular that points of origin are obscured within tens of millions of years by the chaotic nature of orbits in the galactic potential.

Remarks have also given consideration to the idea that `Oumuamua was originally ejected from the Solar System and has \textit{re-encountered} the Sun. This possibility has an exceptionally low probability. The phase mixing \citep{BinneyTremaine2008} implied by `Oumuamua's large heliocentric $v_{\infty}$ implies that its trajectory was fully independent from that of the Sun's prior to its celebrated recent encounter. Adopting a $\sigma=\pi\,{\rm AU}^2$ cross section (that includes the Sun's gravitational focusing), a $\langle v \rangle=25\,{\rm km/s}$ velocity, and solar excursions of $\Delta R=\pm350$~pc and $\Delta Z=\pm350$~pc centered on a $R=8.5$~kpc mean galactocentic radius, one calculates an expected wait time $\tau=(2\pi R\Delta R \Delta Z)/(\sigma \langle v \rangle) \sim 10^{24}$~ years between departure from the Sun's vicinity and its subsequent return.

Although the physical origin of `Oumuamua is speculative, a number of observations provide clues to its composition and layout. At periastron, `Oumuamua passed within $0.25\,{\rm AU}$ of the Sun, and despite achieving solar irradiation levels, $I>20\,{\rm kW\,m^{-2}}$,  deep images of the object produced no sign of a coma \citep{Mpec2017b}, demonstrating that it has a non-volatile near-surface composition. Naively, this seems surprising, since estimates of the ratio of cometary to asteroidal material in the Oort cloud range from 200:1 to 10,000:1 \citep{Meech2017}. Strong color-independent time variability in the photometric light curve observations presented by \citet{Knight2017}, \citet{Jewitt2017}, \citet{Meech2017}, and \citet{Bolin2017}, suggest that the object is highly elongated and moderately rapidly rotating, with an aspect ratio of up to 10:1, a mean radius of $r_{\rm o}=102 \pm 4\, {\rm m}$ and a $P_{\rm rot}\sim 7.3\,{\rm h}$ rotation period. This rate of spin is sufficient to overwhelm the self-gravity of a rubble pile with $\rho \lesssim 6\,{\rm g\,cm^{-3}}$, implying that the `Oumuamua has non-negligible tensile strength. \citet{Fraser2017} and \citet{Drahus2017} independently analyzed `Oumuamua's photometric light curve measurements, and concluded that it is exhibiting non-principle axis rotation. Like the Saturnian satellite Hyperion \citep{Wisdom1984},  `Oumuamua may be tumbling chaotically.

\citet{Domokos2017} propose (drawing on an Eikonal model) that `Oumuamua's highly elongated shape is the expected consequence of abrasion incurred by extended passage through a sea of tiny particles, and \citet{Gaidos2017b} notes that the large aspect ratio could potentially arise if `Oumuamua is a contact binary composed of two less-extremely elongated ellipsoidal components with heterogeneous surfaces.

 `Oumuamua's unusual observed properties generated discussion of a full range of theories regarding its provenance. Ensuing attempts to detect technologically generated radio signals from the asteroid using eight hours of time on the Robert C. Byrd Green Bank Telescope were, however, unsuccessful \citep{Enriquez2018}.

Planet formation theory suggests that giant planet migration may be responsible for the ejection of very large quantities of planetesimals into interstellar space \citep{Tsiganis2005, Levison2008}.  If `Oumuamua is a planetesimal ejected from another system, then the observation of one such comet during Pan-STARRS' operational lifetime to date implies that an average of approximately $1M_\oplus$ of material is ejected from every star \citep{Laughlin2017, Trilling2017}.  \citet{Raymond2017} point out that this estimate assumes a mono-sized population of ISOs, and that if the true planetesimal mass distribution is top heavy, similar estimates yield an implausibly high mass of ejected material for each star. They demonstrate that this may be reconciled if `Oumuamua is a member of a sub-population of  $0.1-1 \%$ of planetesimals that are tidally disrupted by gas giant planets on their pathway to ejection.

The advent of `Oumuamua, which was detected in spite of a number of observational biases, suggests that a large population of similar objects may pervade the galaxy.  Extrapolating from a sample size of one, and using cross section-based estimates, \citet{Laughlin2017} calculate that the Galaxy contains $N\sim2\times10^{26}$ `Oumuamua-like planetesimals. This result concords with estimates by \citet{Jewitt2017}, \citet{Meech2017} and \citet{Trilling2017} who  estimate space densities of order $n\sim0.1\,{\rm AU^{-3}}$, a value implying that on average, of order one such object can always be found traversing an $R=1\,{\rm AU}$ sphere centered on the Sun. Most recently, \citet{Do2018} carried out a detailed assessment of Pan-STARRS' ability to detect `Oumuamua-like objects. Their evaluation folds in detectability as a function of direction of approach, and considers possible power-law indices for the unknown size distribution. \citet{Do2018}  find an interstellar number density, $n=0.2$ of similar objects per $\rm{AU}^{3}$. Clearly, additional detections will be both welcome and informative.

\citet{Jewitt2017} draw analogies  with the composition and the behavior of comets to argue that `Oumuamua's lack of coma is not inconsistent with an ice-rich bulk composition. Long-duration exposure to the radiation environment of the interstellar medium is capable of mantling a comet-like object with a crust of volatile-depleted ``regolith'' of very low thermal conductivity \citep{Cooper2003}. \citet{Fitzsimmons2017} present a more detailed thermal model that supports the discussion in \citet{Jewitt2017}. \citet{Raymond2018b} posit that `Oumuamua is a fragment of a tidally disrupted comet-like, icy planetesimal, that underwent enough close passages to its host star to become extinct. Additionally, \citet{Jewitt2017} raise the interesting possibility that heat transfer into `Oumuamua may cause it to develop an observable coma as it departs the Solar System. A related phenomona has been observed for select comets, including comet P/Halley, which produced a post-perihelion outburst at 14 AU that has been attributed to release of trapped gasses during a phase transition at depth from amorphous to crystalline ice \citep{Prialnik1992}. It is unlikely, but nonetheless possible that HST observations of `Oumuamua, completed in November and December of 2017, and at the beginning of 2018\footnote{\url{http://www.stsci.edu/hst/phase2-public/15405.pdf}}, may show signs of coma, thereby placing important constraints on both the surface and interior properties.

\citet{Cuk2017}, and \citet{Jackson2017}, however, find no particular surprise in observing an ISO with an asteroidal bulk composition. These authors argue that `Oumuamua was ejected by a binary system, in which case it (and many bodies similar to it) may naturally be composed of non-volatile material. \citet{Hansen2017} point to the observed rate of pollution of white dwarfs by asteroidal debris, and argue that accompanying ejection of such material from host systems must also occur. Moreover, they note that large post-main sequence stellar luminosities will lead to volatile loss for many ejected bodies. As a point of reference, the Sun is expected to achieve a peak luminosity of $5200\,L_{\odot}$ \citep{Sackmann1993}, and will loose a significant burden of its minor-body retinue in the dynamical reshuffling that will befall the Solar System as the Sun loses mass during its transition into a white dwarf \citep{Duncan1998}. \citet{Rafikov2018} also considers the hypothesis that `Oumuamua's lack of coma implies a refractory composition, and demonstrates that tidal disruption of asteroidal bodies that make close approaches to white dwarf stars can potentially explain its origin.

Given the range of possible origins, it would be useful to investigate a body such as `Oumuamua in close proximity, especially in connection with the use of a kinetic impactor to excavate a debris plume that could be examined spectroscopically. Projects of this type certainly have precedent. Over the past three decades, a rich heritage of \textit{in situ} investigations of comets, asteroids, and Kuiper Belt Objects has been established, and in particular, a number of missions have been targeted to Solar System comets. The first missions of this type included the NASA International Sun-Earth Explorer (ISEE-3) which was directed through the tail of Comet Giacobini-Zinner in 1985 \citep{Tsurutani1986}, and the Vega 1, Vega 2, Sakigake, Suisei, and Giotto missions that all investigated Comet Halley in 1986 \citep{LePage2011}. 

NASA's Deep Space 1 Mission encountered Comet Borrelly in 2001 \citep{Nelson2004}. In 2004, NASA's Stardust Mission flew to within 240 km of the nucleus of comet Wild 2 \citep{Tsou2004}, collected dust particles by both trapping them in aerogel and cratering them on foil, and returned the payload to Earth in 2006 \citep{Brownlee2006, Flynn2006, Sanford2006}. During the near-contemporaneous Deep Impact Mission, a purpose-built kinetic impactor was directed to strike the comet Tempel 1, producing a debris plume that was observed both from Earth and from the Mission's main spacecraft. The impact excavated contents from beneath the comet's radiation-hardened surface, allowing an assessment of pristine planetesimal-forming material \citep{Ahearn2005}.

More recently, ESA's Rosetta Mission \citep{Glassmeier2007} deployed a small lander on the surface of comet 67/Churyumov-Gerasimenko, while the main Rosetta craft orbited at a distance of $d \sim\,$20 km and extensively imaged the comet's surface. At mission's end in 2016, the main craft was directed to strike the comet, returning extremely close-up images as it did so \citep{El-Maarry2015}. Finally, although not directed to visit a comet, NASA's Osiris-REx mission is flying to the asteroid Bennu, where it will sample the surface and return the contents to Earth \citep{Lauretta2012}.

The inference that extra-solar planetesimals frequently infiltrate the inner Solar System raises the possibility of designing a current-technology space mission to intercept and investigate one of these objects at close range. Such a mission would face a number of challenges, including (1) the large heliocentric velocities of objects on hyperbolic trajectories, and (2) the lack of substantial time following the discovery of the target object for mission planning and execution, and (3) uncertainty in targeting during final approach. It is worth noting that when `Oumuamua was detected and announced in late October 2017, it had already passed its periastron location (which occurred on 9 September, 2017), and indeed, was already more than 1 AU from the Sun.

\citet{Hein2017} reported a preliminary investigation of mission designs aimed at intercepting `Oumuamua itself. At present (and especially going forward) such missions are very challenging to mount, due to the large $\Delta v$'s required to catch up with the rapidly departing object. \citet{Hein2017} considered trajectories that include direct transfers, as well as missions that employ a combination of Jovian gravity assist and Solar Oberth maneuvers \citep[see, e.g.][]{Bond1996}, and they speculated on the potential use of advanced technologies such as solar sails, laser propulsion, and magnetospheric ``chipsat'' acceleration. 

In this paper, we adopt a complimentary approach and assess the near-term prospects for intercepting future `Oumuamua-like arrivals using conventional chemical propulsion. We envision a mission with scientific charge that is similar to the Deep Impact Mission, in which a kinetic impactor strikes an impinging ISO at high velocity, enabling a companion flyby probe to carry out a spectroscopic examination of its surface and crucially, its sub-surface contents. The feasibility of such missions hinges primarily on whether an incoming body can be detected with sufficient lead-time to launch an interceptor that can meet an ISO traveling on a hyperbolic trajectory. An evaluation of this problem thus involves coupling detectability analyses \citep{Moro2009, Cook2016} to trajectory evaluations for a population of incoming bodies.

The plan for this paper is as follows. In \S 2, we outline a thermal model for `Oumuamua during its passage through the inner Solar System, allowing us to place quantitative limits on the composition and the thermal conductivity of its surface.  This analysis delineates that the primary scientific goal for any mission to such a body would be to determine whether the regolith-sheathed ice model is correct; in augmentation of efforts by other authors, the model that we consider takes both the elongated shape, the full mass, and the chaotic tumbling of the body into account. In \S 3, we assess the likelihood of arrival directions of future ISOs as a function of arrival direction (based on a heliocentric viewpoint), and we predict detection rates as a function of observing season (on Earth) and limiting survey magnitude. In \S 4 we evaluate the expected fuel (or specifically, $\Delta v$) requirements for short-notice direct-transfer interception missions launched from Earth to investigate arriving ISOs. We pay particular attention to the range of trajectories that could have been used to intercept `Oumuamua (and find that they could easily have been mounted if the ISO had been detected in time). In \S 5, we enumerate the potential scientific benefits of high relative velocity intercept missions and conclude.

\section{`Oumuamua's Thermal Response to Solar Warming}

`Oumuamua's unusually elongated shape and its lack of a coma were both unexpected, yet these two features may also be connected. The highly elongated figure of the object, coupled with its tumbling (that is, non-principal axis rotational) motion, imply that during the multi-month journey through the inner Solar System, the surface was repeatedly and randomly illuminated from all directions.  Depending on the viewing angle, `Oumuamua's apparently spindle-like geometry allows  the ratio, $\xi$, of the projected area, $A_p(\theta,\phi)$ to the surface area, $S_A$, to instantaneously be substantially smaller or larger than the value, $\xi=1/4$ appropriate to a sphere.

Moreover, as discussed in the previous section, it is reasonable to expect that `Oumuamua is covered with a porous regolith of material with low thermal conductivity that has been significantly processed during billions of years of exposure to both the galactic cosmic ray flux and to penetrating ionizing radiation. Given the observed lack of coma, and adopting an ellipsoidal three-dimensional shape, we can estimate the combinations of regolith thickness and thermal diffusivity required to protect a frozen, volatile-rich interior from sublimation. Our estimate proceeds by first outlining the calculation of $\xi$ for an adopted body shape and a choice of orientation (or ranges of orientation), and then solving for the rate of heat transfer through the regolith given (i) the solar illumination during the known orbital trajectory, (ii) an appropriately time-averaged form factor, $\xi$, and (iii) a parameterized grid of assumptions for the regolith depth and thermal conductivity.

\subsection{Geometry of a Projected Ellipsoid}

A highly elongated object displays a relatively small projected area relative to its surface area when viewed along its long axis, and a similarly large one when viewed along its short axis. This may have served to limit the solar radiation energy that `Oumuamua's surface received during its closest approach to the Sun. For quantitative heat transfer calculations, we need an analytic framework to estimate the value of the geometric factor, $\xi$, as a function of orientation.

 In general, a two-dimensional ellipse of the form\footnote{Note that the $2a_{12}xy$ term rotates the orientation of the ellipse about the origin. We neglect first degree polynomial terms, which produce translations.}

\begin{equation}
    a_{11}x^2+2a_{12}xy+a_{22}y^2+c = 0 \, ,
\end{equation}
where $x$ and $y$ denote the spatial coordinates and $a_{11}$, $a_{12}$, $a_{22}$ and $c$ are general coefficients, has an area, $A$,
\begin{equation}
    A=\frac{\pi}{\sqrt{\lambda_+\lambda_-}}\, ,
\end{equation}
where $\lambda_+$ and $\lambda_-$ are the eigenvalues of the general quadratic form for the ellipse equation, 

\begin{equation}
    \lambda_{+,-} = \frac{1}{2c} [-(a_{11}+a_{22})\pm\sqrt{(a_{11}-a_{22})^2+4a_{12}^2}] \, . 
\end{equation}

If the tumbling time scale (which is of order the rotation period) is short in comparison to the encounter time scale, $\xi$ is simply the  projected surface area of an ellipsoid, averaged over all viewing angles.  Cauchy proved that for any convex surface, this is always $\frac14$. Simply stated, for the sphere,  $\xi=1/4$, and because of the symmetry of the sphere, this result holds for each infinitesimal surface area element on the sphere. Therefore, any infinitesimal element of area has  $\xi=1/4$. Since the surface of any convex body consists of a number of these infinitesimal surface area elements, Cauchy's theorem follows by summing all of these contributions \citep{Meltzer1949}.

If however, `Oumuamua's orientation was predominantly head on or face on during its periastron passage (when the received flux was changing rapidly), this could serve to significantly increase or decrease the amount of solar radiation that it was exposed to.

Adopting the notation of \citet{Stark1977}, who derived a similar quantity for the isophotes of elliptical galaxies, we consider an ellipsoid centered in a 3-dimensional coordinate system $(x,y,z)$  with coefficients $t$, $u$ and $a_v$, of the form
\begin{equation}\label{eq:Ellipsoid}
    (tx)^2+(uy)^2+z^2=a_v^2 \, .
\end{equation}

A rotation through the first two Euler angles is sufficient to transform to any viewing angle, so consider a rotation of $(\theta,\phi)$,
\begin{equation}
    \left[ \begin{array}{c} x \\ y \\ z \end{array} \right] = 
    \begin{bmatrix} \cos{\phi} & -\cos{\theta}\sin{\phi} &\sin{\theta}\sin{\phi}  \\ \sin{\phi} & \cos{\theta}\cos{\phi} &-\sin{\theta}\cos{\phi} \\ 0&\sin{\theta}&\cos{\theta} \end{bmatrix}  \left[ \begin{array}{c} x'\\ y' \\ z' \end{array} \right] \, .
\end{equation}

Then, in the rotated coordinate system, $(x',y',z')$, Equation \ref{eq:Ellipsoid} is,
\begin{equation}\label{eq:RotatedEllipsoid}
    f z'^2+gz'^2+h = a_v^2 \, ,
\end{equation}
where
\begin{equation}
    f = f(\phi,\theta,t,u) = t^2\sin^2{\theta}\sin^2{\phi}+u^2\sin^2{\theta}\cos^2\phi+\cos^2\theta \, ,
\end{equation}
\begin{equation}
\begin{split}
 g = g(x',y',\phi,\theta,t,u) = \sin\theta\sin{2\phi}(t^2-u^2)x'+\\ \sin{2\theta}(1-t^2\sin^2\phi-u^2\cos^2\phi)y' \, ,
\end{split}
\end{equation}
and
\begin{equation}
\begin{split}
     h = h(x',y',\phi,\theta,t,u) = (t^2\cos^2\phi+u^2\sin^2\phi)x'^2+\\\cos\theta\sin{2\phi}(u^2-t^2)x'y'+\\(t^2\cos^2\theta\sin^2\phi+u^2\cos^2\theta\cos^2\phi+\sin^2\theta)y'^2 \, . 
\end{split}
\end{equation}
 Therefore, at a given viewing angle, $(\theta,\phi)$, the 2-dimensional projection of the ellipse is defined by the maxima of the rotated coordinates orthogonal to the line of sight along the ellipsoid. For a rotation through the first Euler angle $\phi$, with line of sight defined by $x'$, the projected ellipsoid has a semi-major axis along $y'$, defined by the maxima of $y'$, where $\frac{\partial y'}{\partial x}=0$, and semi-minor axis along the $z'$ axis. 

The area of this projected ellipse, $A_p$, is then
\begin{equation}\label{eq:Aproj}
\begin{split}
    A_p=\frac{\pi}{tu}\bigg(\frac{u^4\sin^2{\phi}+t^4\cos^2{\phi}}{t^2\cos^2{\phi}+u^2\sin^2{\phi}} \bigg)^{\frac12}\, .
\end{split}
\end{equation}

For a triaxial ellipsoid of aspect ratio 10:1:1, the value reported by \citet{Meech2017}, if viewed head-on, the form factor is minimized with $\xi \sim0.03$, whereas if the body is viewed along the short axis, one finds a maximum value, $\xi \sim 0.31$. 

\begin{figure*}
\begin{center}
\resizebox{0.99\textwidth}{!}{\includegraphics*[trim={.01cm .15cm .00cm .04cm},clip]{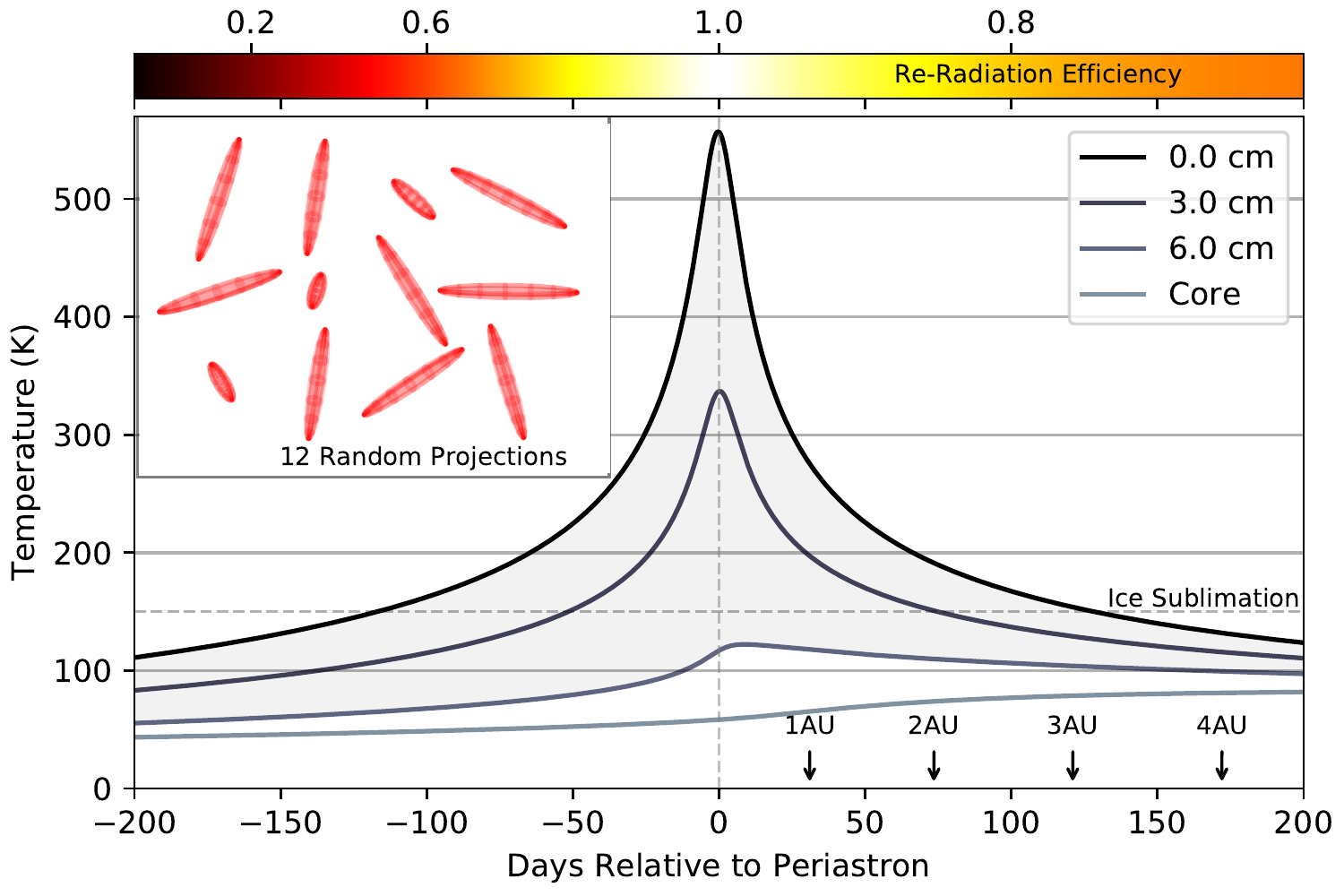}}
\caption{Thermal modeling of `Oumuamua during its solar encounter. We show the temperature profile at depths of 0.0, 3.0, 6.0 cm (the regolith-ice interface) and at the core (at 900cm) over the 200 days prior to and after periastron. The upper left inset shows twelve different projections of a 10:1:1 triaxial ellipsoid defined by random rotations though the first two Euler angles, to provide a visual depiction of the substantial variation in projected surface area, due to `Oumuamua's chaotic tumbling (see Eqn. \ref{eq:Aproj}). The top panel shows the efficiency of blackbody radiation during the flyby, evaluated as the instantaneous energy radiated/energy received. In the simulation, the icy asteroid is coated in a 6 cm thick layer of porous regolith material, as indicated by the shading, and we use thermal properties typical of such materials (numerical values of which are cited in section 2.2). We verify that globally, energy in the simulation is conserved to better than a factor of $\sim1\%$ of the cumulative sum of the total energy received minus the total radiated, using a Bond albedo of $\alpha=.01$ and bolometric emissivity of $\epsilon=.95$, as presented in similar simulations by \citet{Fitzsimmons2017}. The arrows indicate the times where `Oumuamua reaches distances of 1, 2, 3 and 4 AU from the Sun as it exits the Solar System. With this choice of depth and conductivity, `Oumuamua never produces a visible coma.} 
\label{fig:ThermalProfile}
\end{center}
\end{figure*}

\begin{figure}
\begin{center}
\resizebox{0.49\textwidth}{!}{\includegraphics*[trim={.10cm .15cm .0cm .8cm},clip]{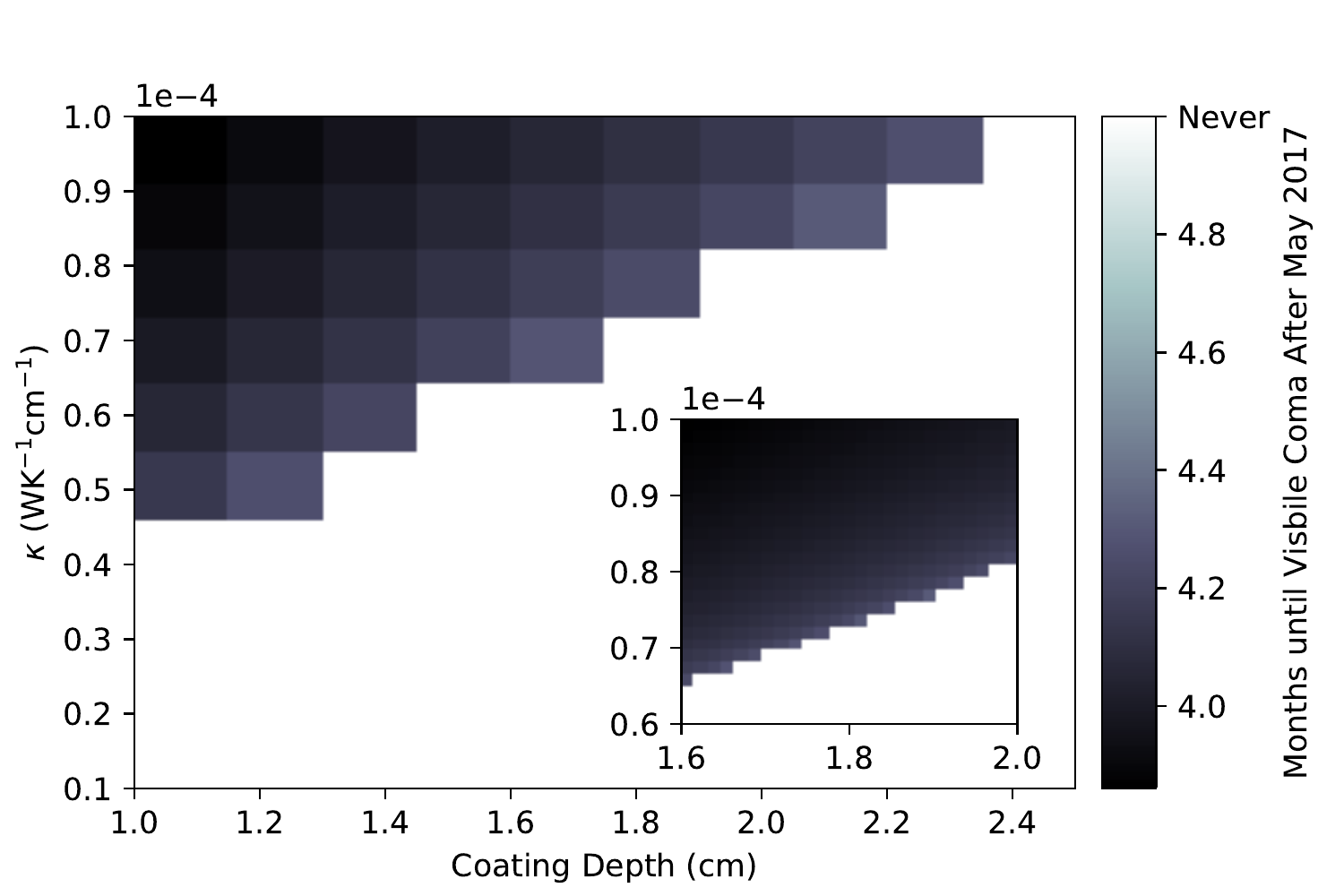}}
\caption{The expected appearance date of a coma as a function of thermal conductivty $\kappa$ and depth of the insulating coating around `Oumuamua. The white coloring represents model parameters that never produce a coma. The simulations presented in the main panel were performed on  a 10x10 grid of thermal conductivity and depth of the coating, using $\xi\sim0.096$. We present a finer 25x25 grid of simulations on an interesting subset of the larger grid, presented in the bottom right panel with the same color scale.  Region of parameter space predicting a visible coma  (delineated by the dark-to-light border) can be ruled out.} 
\label{fig:kappadepth}
\end{center}
\end{figure}

\subsection{`Oumuamua's Thermal Response to Solar Heating}
Assuming that the timescale for randomization of the principal axis orientation from tumbling, $\tau_{ran}\sim\tau_{rot}$, is shorter than the time over which the Sun to `Oumuamua distance changes significantly, we can adopt the approximation $\xi=1/4$ that follows from Cauchy's theorem and calculate the solar flux, $\Phi$, that `Oumuamua received at any point during its trajectory,
\begin{equation}\label{eq:Flux}
    \Phi(t) = \frac{L_\odot }{16\pi d(t)^2}(1-\alpha) \, ,
\end{equation}
where $d$ is the instantaneous solar distance,  $L_\odot$ is the solar luminosity, and $\alpha$ is the surface Bond albedo.

To model the evolution of the temperature profile of `Oumuamua during its encounter with the Sun, we solve the one-dimensional heat conduction equation in cylindrical coordinates,

\begin{equation}\label{eq:CylHeatEq}
    \frac{dT}{dt}=\frac{\kappa}{\rho c_P}\frac{1}{r}\frac{\partial}{\partial r}\big(r\frac{\partial T}{\partial r}\big) \, ,
\end{equation}
for the temperature $T$, where $\kappa$ is the thermal conductivity, $\rho$ is the density, $c_P$ is the specific heat and $r$ is the radial depth.

We solve the above equation on a 200 zone grid, to model the entirety of the 1800 cm semi-minor axis of the best fit ellipsioid from \citet{Meech2017}. The model has an outer shell of low-conductivity ``regolith'' and an inner core of $\rm{H_2 O}$ ice. For computational efficiency, we use ice zones that are substantially larger than the regolith zones. 

To enforce global energy conservation, we adopt boundary treatments for (i) the outer-most regolith surface zone, (ii) the  regolith-ice transition interface, and (iii) the inner-most ice zone. For the surface regolith zone, we use a Newton-Raphson technique to find a surface temperature, $T$ that balances absorption of solar flux, heating of the surface zone material, reradiation from the surface, and diffusion of heat energy from the surface zone into the interior,

\begin{equation}\label{eq:OuterBC}
    \Phi(t) -\epsilon \sigma T^4 - \kappa  \frac{dT}{dr}_{r=0}-c_P \rho \Delta r \Delta T=0\, .
\end{equation}
 In equation \ref{eq:OuterBC}, $\Phi(t)$ is the energy flux received in the outer-zone, $\epsilon \sigma T^4$ is the energy radiated, $\kappa\, \,{dT/dr}_{r=0}$ is the energy diffused into the adjoining zone, and $c_P \rho \Delta r \Delta T$ is the internal energy change of the zone resulting from the change in temperature. We use $\epsilon \sim.95$ for the bolometric emissivity, and $\sigma$ is the Stefan-Boltzmann constant.  At the ice-regolith interface and the inner boundary, we also use Newton-Raphson iteration to solve similar equations, albeit without the radiative energy terms. For the regolith-to-ice transition condition, we consider both the flux diffused from the inner-most regolith zone to the first ice zone, and from the first ice zone to the second ice zone.

Figure \ref{fig:ThermalProfile} shows the results of this thermal modeling applied to `Oumuamua during solar encounter. The temperature profile is plotted at depths of 0.0, 3.0, 6.0 cm and at the center over the 200 days prior to and after periastron. The full simulation was run over a multi-year period spanning 2013/02/09 to 2019/12/09. We use a 200-zone grid representing the asteroid  coated in a  6 cm thick layer of porous regolith material. We use  values of density $\rho_r=1.0\,{\rm g\,cm}^{-3}$ and $\rho_{ice}=0.92\,{\rm g\,cm}^{-3}$,  specific heat capacities $c_{Pr}=0.55\,{\rm J\,g}^{-1}{\rm K}^{-1}$ and $c_{P{\rm ice}}=2.0\,{\rm J\,g^{-1}K^{-1}}$, and thermal conductivities $\kappa_r=9.0\times10^{-5}\,{\rm W\,K^{-1}\,cm^{-1}}$ $\kappa_{\rm ice}=2.25\times10^{-2}\,{\rm W\,K^{-1}cm^{-1}}$ for the regolith and ice respectively. We track the energy flux recieved, radiated and contained in the simulation, and chart the re-radiation efficiency in the top panel of Figure \ref{fig:ThermalProfile}. We require that the total energy  is conserved to a factor of better than $\sim1\%$ of the sum of the total energy received minus radiated.  With this choice of depth and conductivities, a visible coma is absent at all times.

It is worthwhile to discuss the apparent differences between our results and those of \citet{Jewitt2017} and \citet{Fitzsimmons2017}. \citet{Jewitt2017} make an order of magnitude estimation based on the heat equation, $t_c\sim d^2\rho c_P/\kappa$, to estimate the thermal skin depth $d$ (at which the surface temperature is a factor of $\sim 1/e$ times the surface temperature $T\lesssim 560K$) of the regolith coating, where $t_c$ is the timescale to conduct heat. Using thermal properties similar to those assumed in this simulation and $t_c =2\times10^7$s, they estimate that `Oumuamua has a skin depth $d\sim 50\,$cm, and conclude that to prevent a visible coma, the depth of the thermal coating must be $D \gtrsim 50\,$cm. While this provides a first estimate for the coating's thickness, it does not account for the geometric factor $\xi$ from the smaller projected surface area, the efficiency of thermal re-radiation over the course of the flyby, and the potentially sharp difference in thermal conductivities at the (assumed discontinuous) ice-regolith interface. We find that with $\xi\sim .25$  and a radiation efficiency that approaches unity close to and after periastron, the surface temperature heats to $T\lesssim550K$. We also see that the presence of a large reservoir of high-conductivity ice beneath the regolith permits a much steeper gradient in regolith temperature due to the ice-regolith boundary condition. Additionally, a slight decrease in temperature diffusion stems from the adoption of cylindrical coordinates. 

\citet{Fitzsimmons2017} present high-resolution simulations of the thermal regolith coating down to 5 meters in depth. This analysis includes the thermal radiation, and uses a geometric factor of $\xi=1/\pi$, a regolith density of $\rho_r=1.0\,{\rm g\,cm}^{-3}$,  specific heat capacity of $c_{Pr}=0.55\,{\rm J\,g}^{-1}{\rm K}^{-1}$ and thermal conductivity of $\kappa_r=1.0\times10^{-5}\,{\rm W\,K^{-1}\,cm^{-1}}$. In their simulations, the surface temperature heats to $T\lesssim580K$, and the temperature is reduced by a factor of $1/e$ at a depth of $\sim10-20\,$cm. Using the order-of-magnitude argument from \citet{Jewitt2017} applied to the thermal properties assumed in the \citet{Fitzsimmons2017} results, one would expect the skin depth to be $d\sim 15\,$cm which is consistent with their simulation. The authors note that the skin depth would scale with different thermal properties and albedos than those chosen in their simulation. We verify that our simulation converges to the \citet{Fitzsimmons2017} results with the same thermal and geometric properties and an increased amount of regolith coating. 

The major discrepancy in between our analysis and that of \citet{Fitzsimmons2017} is the smaller effective skin depth of the regolith, which can be predominantly explained by the additional modeling of the ice. With the thermal properties assumed in Figure \ref{fig:ThermalProfile}, the order-of-magnitude estimate of the regolith skin depth is $d\sim 40\,$cm,  which is inconsistent with our simulation. The timescale for heat diffusion over a constant distance scales as $t_c\sim \rho c_P/\kappa$. This implies that with the values used by \citet{Fitzsimmons2017}, the timescale for heat diffusion of ice is $t_{cice}\sim t_{cr}/100$. With these choices of thermal properties, heat diffused across the ice-regolith boundary effectively heats up the entire interior of the icy asteroid, maintaining a substantially colder inner regolith temperature, very close to the temperature of the core. With the thermal properties used in Figure \ref{fig:ThermalProfile}, the timescale for heat diffusion of ice is $t_{cice}\sim t_{cr}/10$. With the extra factor of $10$ in the diffusion timescale, a temperature gradient is now maintained in the ice zones, evident in the difference in the core and  6 cm temperature profiles in Figure \ref{fig:ThermalProfile}, but the first regolith zone is still substantially colder than it would be without the ice. Therefore, equating the skin depth of regolith to the depth of coating at which the icy interior will sublimate is potentially misleading, as the regolith will exhibit a much higher temperature gradient due to ice-regolith boundary condition.  We have verified that the results of these simulations have converged, and that the discrepancy is not due to the resolution. In summary, we conclude that the discrepancies can be attributed to the geometric factor, thermal radiation and the modeling of the icy interior. We note that if $\xi>.25$ or the asteroid was smaller and had less ice to heat up, more regolith coating would be required to prevent development of a coma.

We present similar simulations for a range of depth and thermal conductivities of the regolith coating, and the results are presented in Figure \ref{fig:kappadepth}. For our grid, we choose a range of thermal conductivities $\kappa_r \in (1.0\times10^{-5},1.0\times10^{-4})\,{\rm W\,K^{-1}\,cm^{-1}}$ that allow for a factor of 10 difference in the timescale for diffusion across the regolith. At the minimum value of $\kappa_r$, the temperature gradient in the ice is very small, leading to a much colder inner regolith zone, while at the maximum $\kappa_r$, a temperature gradient is maintained in the ice. We see that generally, a coma is only produced very close to periastron, with a very small amount of insular coating.   We conclude that future ISOs may similarly fail to produce comas in response to solar irradiation. As a consequence, the most definitive way to determine the bulk composition of ISOs may be through excavation of subsurface material using a kinetic impactor.

\section{Trajectory Expectations for Future Interstellar Asteroids}

If estimates of high space densities for `Oumuamua-like objects prove correct, discoveries of similar objects will certainly be forthcoming, and it is useful to assess the expected distributions of arrival directions, speeds, and impact parameters with respect to both the Sun and the Earth.

\begin{figure}
\begin{center}
\resizebox{0.49\textwidth}{!}{\includegraphics*[trim={1.830cm 1.55cm .65cm 1.5cm},clip]{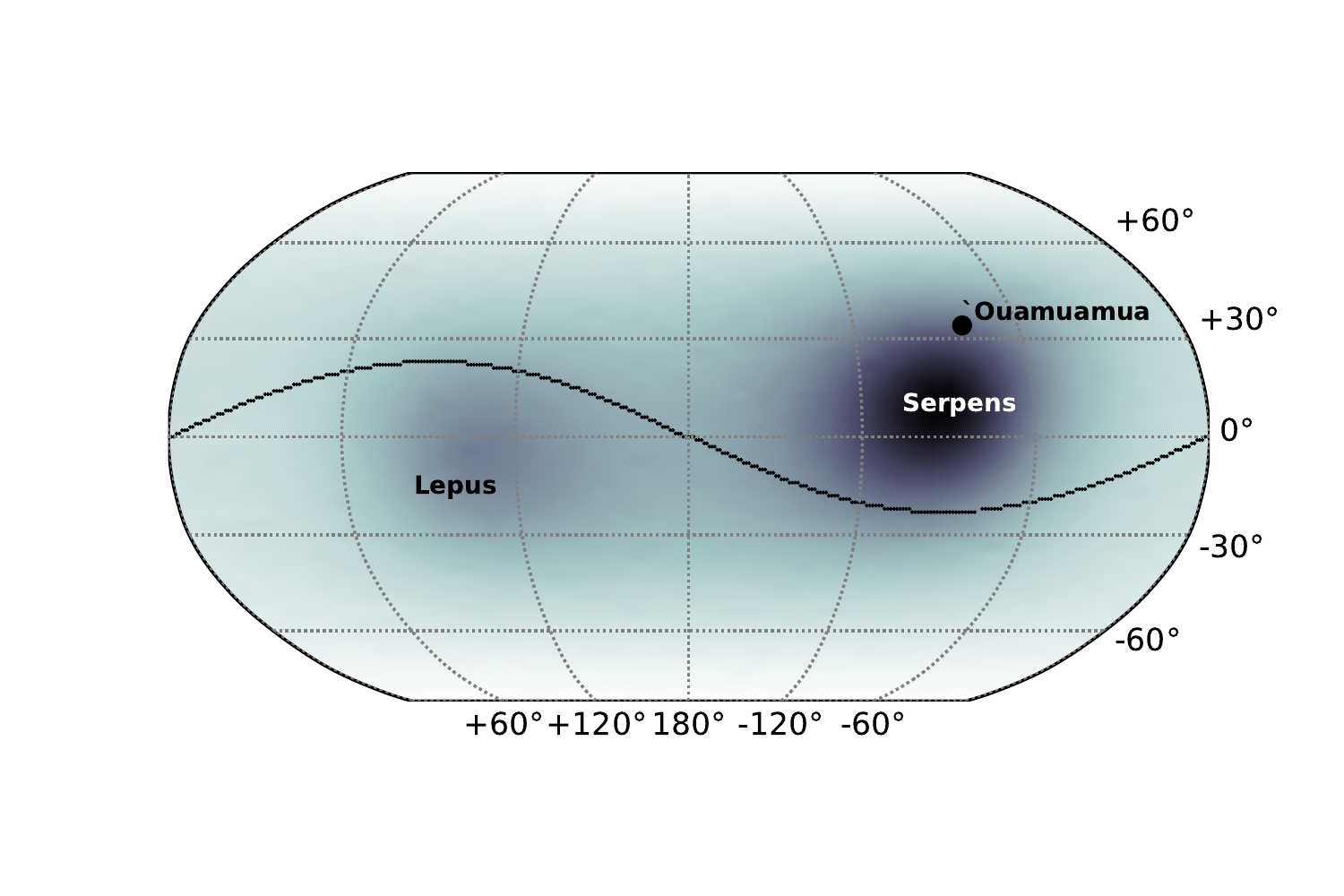}}
\caption{A sky map showing the probability that a future interstellar asteroid will approach the Solar System on a trajectory parallel to that direction. The darker colors indicate a higher probability.  The axes denote degrees from a heliocentric point of view and the ecliptic is plotted in black.  The sky positions of the constellations Serpens and Lepus, which are close in proximity to the Solar apex and anti-apex respectively, are plotted for context.  The black circle indicates the sky location that `Oumuamua entered our Solar System, consistent with the prediction that the majority of these objects will approach with velocities parallel to the galactic apex.}
\label{fig:Skymap}
\end{center}
\end{figure}

\begin{figure}
\begin{center}
\resizebox{0.49\textwidth}{!}{\includegraphics*[trim={.30cm .15cm .85cm .8cm},clip]{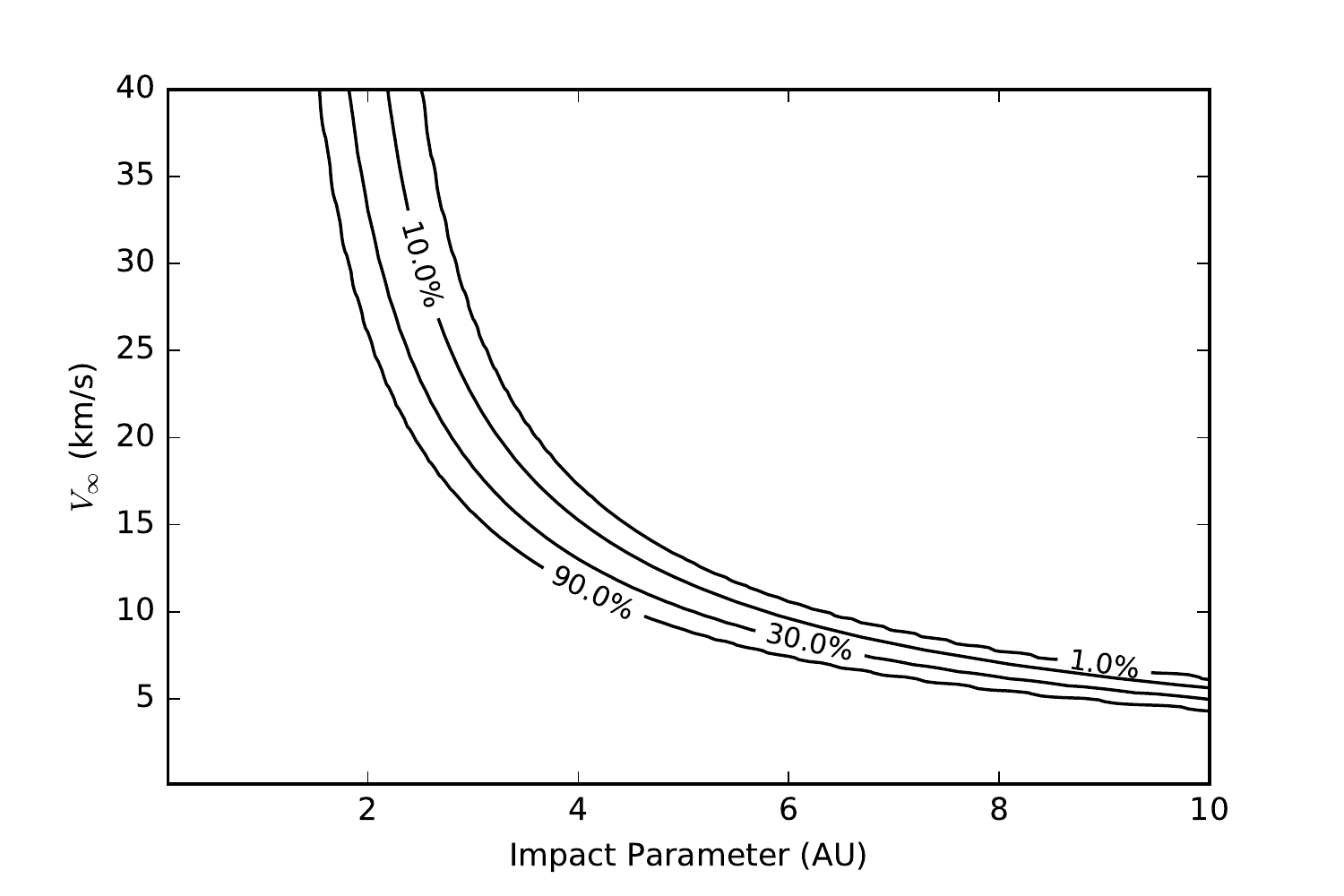}}
\caption{Contours depicting the percent of `Oumuamua-like ISOs that will reach an apparent magnitude $m\le24$ before periastron as a function of excess velocity and impact parameter. On a 100x100 grid of $b$ and $v_\infty$, we calculate the asteroid's two-dimensional trajectory, and  consider all possible positions of the Earth to be on the 1 AU unit sphere. We calculate the fractional surface area on the sphere where $m\le24$ at any point before periastron.}
\label{fig:detections}
\end{center}
\end{figure}

 If we assume that the population of ISOs in the galaxy has a kinematic distribution similar to Population I stars, we can adopt the phase space distribution function for the stellar population in the solar neighborhood \citep{BinneyTremaine2008} as a proxy for that of the free-streaming interstellar asteroids. Given that the the number density of `Oumuamua-like interstellar asteroids may be as much as $\sim10^{16}$ higher than the stellar number density, the collisionless fluid approximation is even more applicable to this population.

We assume that the interstellar asteroids in the Solar System's vicinity follow a Schwarzschild velocity distribution, where the dispersions, $\sigma_1$, $\sigma_2$ and $\sigma_3$, come from the Hipparcos catalogue \citep{BinneyTremaine2008}. The coordinate system with respect to a right-handed galactic coordinate is given by the vertex deviation for the primary eigenvector of the ellipsoidal distribution function for a given population of stars. The number density of interstellar asteroids, $n({\bf \Delta v})$,  in a volume of velocity space ${\bf \Delta v} = (\Delta v_1,\Delta v_2, \Delta v_3)$, is

\begin{equation}\label{eq:Schwarzschild}
    n({\bf \Delta v}) =\frac{n}{8} \prod_{i=1}^3 \textrm{erf}(\frac{v_i}{\sqrt{2}\sigma_i})|_{v_i-\Delta v_i}^{v_i+\Delta v_i}  \, .
\end{equation}

In the vicinity of the Sun, gravitational focusing increases the flux of arriving bodies, especially for periastron distances, $r<1$ AU. To quantify this effect, consider a test particle on a hyperbolic trajectory

\begin{equation}\label{eq:hyporb}
    \frac{1}{r}=\frac{GM_\odot}{b^2v_\infty^2}\big(1+e\cos(\theta-\varpi)\big) \, ,
\end{equation}
 where $(r,\theta)$ are the polar coordinates with respect to the sun's position, $v_\infty$ is the hyperbolic excess velocity, $e$ is the eccentricity, $G$ is the gravitational constant, $M_\odot$ is a solar mass, $\varpi$ is a phase, and $b$ is the impact parameter. In the absence of the Sun's gravitational field, bodies that  enter the 1 AU sphere will have $b\le 1$AU. Gravitational focusing augments the effective cross section by a factor, $\zeta=b^2/1\textrm{AU}^2$, which is a function of  $v_\infty$ for hyperbolic asteroids. To calculate $\zeta$, we find the maximum $b(v_\infty)$ with a periastron distance of $1$AU. Extremizing Equation \ref{eq:hyporb} gives the perihelion distance $r_{p}$, 

\begin{equation}
    r_{p}=\frac{b^2v_\infty^2}{GM_\odot(1+e)} \, .
\end{equation}

Substituting and cancelling gives an expression for the impact parameter $b$. For the case of the Solar System, with numerical values for $r_{p}=1\,{\rm AU}$, $M=1M_{\odot}$, the augmentation for the cross section is, 

\begin{equation}\label{eq:Zeta}
    \zeta \approx1+\big(\frac{42km/s}{v_\infty+\phi}\big)^2 \, ,
\end{equation}
where $\phi$ is an artificial gravitational softening parameter set to $2$ km/s, in order to to avoid the limit of $v_\infty=0$.  We have verified that varying this parameter does not significantly alter the results of this analysis.

We want to connect the velocity distribution in the Sun's galactic neighborhood to incoming ISO trajectories. Every point on the sky isomorphically defines a unit vector with respect to the sun. Likewise, any one of these unit vectors can be mapped to a three-dimensional galactic velocity vector, unique up to a scaling of the associated speed. We construct a three-dimensional grid of galactic velocities, map each point in the grid to its unit vector and corresponding position on the sky, and calculate the associated probability with Equations \ref{eq:Schwarzschild} and \ref{eq:Zeta}. We bin the resulting  grid by  position to produce a full sky probability distribution,  shown in Figure \ref{fig:Skymap}. This map does \textit{not} present the probability that an ISO will perforate the 1AU sphere at that location in the sky. Rather, it  presents the probability that one will approach the Solar System with a velocity vector parallel to that direction, with an arbitrary impact parameter. 

Given a trajectory, the apparent magnitude, $m$, of an object with absolute magnitude $H$ can be calculated using

\begin{equation}\label{eq:magnitude}
    m=H+2.5\log_{10}\big(\frac{d_{BS}^2d_{BO}^2}{p(\Theta)d_{OS}^4}\big)\, ,
\end{equation}
where $d_{BS}$, $d_{BO}$ and $d_{OS}$ are the distances between the body and sun, body and observer and observer and sun, respectively, $\Theta$ is the phase angle defined by

\begin{equation}\label{eq:phaseangle}
    \cos{\Theta} =\frac{d_{BO}^2+d_{BS}^2-d_{OS}^2}{2d_{BO}d_{BS}} \, ,
\end{equation}
and $p(\Theta)$ is the phase integral, approximated by

\begin{equation}\label{eq:phaseintegral}
    p(\Theta) = \frac23\big( (1-\frac{\Theta}{\pi})\cos{\Theta}+\frac{1}{\pi}\sin{\Theta}\big)\, . 
\end{equation}
For given choices of $b$ and $v_\infty$ of an `Oumuamua-like asteroid assumed to have $H=22.4$, we calculate the two-dimensional trajectory, and evaluate the visibility at all points on the 1 AU unit sphere, which (in the approximation of a circular orbit) represents every possible location of the Earth. On a 100x100 grid of $b$ and $v_\infty$, we calculate the fractional surface area on the sphere for which the object is detectable by LSST (m$\le 24$) at any point before periastron, and show contours of the calculated recovery fraction in Figure \ref{fig:detections}.

In order to assess the future detectability and arrival rates of ISOs, we perform two independent Monte Carlo integrations over all possible incoming trajectories. The first integration is computationally efficient because we reduce the problem to a two-dimensional integral via the method described in the previous paragraph. We randomly draw velocities from the Schwarzschild distribution function (\textit{without} the increased gravitational focusing), and impact parameters from $(0,10)$ AU given by a $p(b)\sim b^2$. Given these two parameter choices, we calculate the percentage of objects that are visible from at least one location on the 1AU unit sphere. We do this for 10,000 draws, and see that the solution converges to  $\sim1.6\%$ of incoming ISOs in the Sun-centered $10\,{\rm AU}$ sphere will be detectable. 

In order to estimate wait times and seasonal variability in detectability, we perform a full six-dimensional Monte Carlo integration. We randomly draw a position in the sky given by the distribution in Figure \ref{fig:Skymap},  $v_\infty$ from the Schwarzschild distribution, $b$ from $(0,5)$ AU given by a $p(b)\sim b^2$ distribution, an angle $\psi \in (0,2\pi)$ representing a direction for $b$, and a starting epoch placing Earth at a particular point on its orbit. We position the asteroid in space by first moving to the point 5 AU along the unit vector corresponding to the point on the sky selected, and then translating within the plane normal to this vector by $b$ in a direction defined by $\psi$. (Note that in this setup, the maximum starting distance from the Sun is $5\sqrt{2}$ AU.) We boost the starting velocity to account for the change in gravitational potential energy at the initial position relative to that at infinity. We then use the N-body package Rebound \citep{Rein2012} to integrate both the Earth and the asteroid for 3 years, and evaluate the magnitude of the ISO using Equations \ref{eq:magnitude}-\ref{eq:phaseintegral}  along its trajectory. With a sample size of one, and no observational constraints on the intrinsic magnitude distribution of the galactic population of ISOs, we assume that every ISO has an absolute magnitude of $22.4$. We note, however, that due to Equation \ref{eq:magnitude}, the results of the simulation scale linearly with the absolute magnitude of the ISO. We perform the integration over 3.65 million  trajectories. 

\begin{figure}
\begin{center}
\resizebox{0.49\textwidth}{!}{\includegraphics*[trim={.30cm .15cm .0085cm .008cm},clip]{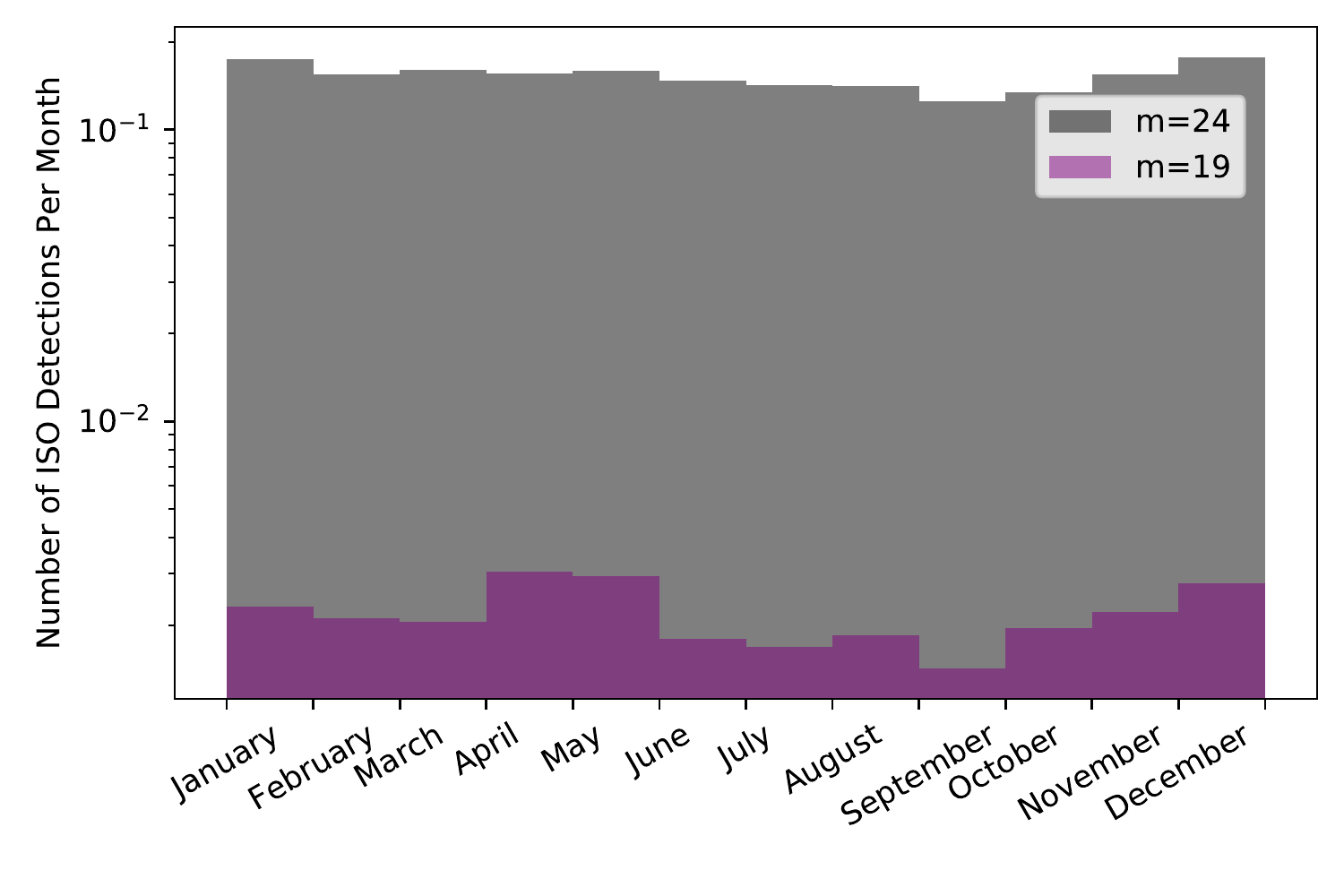}}
\caption{The number of ISOs that first reach apparent magnitudes of $m\le19$ (purple) and $m\le24$ (grey)  per month, under the constraint that the sun, earth, ISO angle be less than 120$^\circ$. We perform a 6-dimensional Monte Carlo integration over possible ISO trajectories, by randomly drawing  sky positions, $v_\infty$, impact parameter $b$, rotation $\psi$  and starting epoch for the Earth. We find that $0.9\%$ of these ISOs will reach  $m\le24$, and $.013\%$ will reach a  $m\le19$.  }
\label{fig:seasonalvariation}
\end{center}
\end{figure}

This method samples the entire distribution of asteroids in the $10$AU cube surrounding the sun, and accurately pinpoints when and where and for how long these objects will be detectable. As a result of our Monte Carlo integration, we find that $2.6\%$ of the ISOs considered during the course of the Monte Carlo simulation will reach  $m\le24$, and $0.017\%$ will reach  $m\le19$.  Additionally, we require that the angle between the sun, earth and ISO be greater than 120$^\circ$ to be detectable by LSST. With this constraint we find that  $0.9\%$ of the ISOs will reach  $m\le24$, and $0.013\%$ will reach  $m\le19$. A number density of $.2$ AU$^{-3}$ \citep{Do2018} implies that ${\bf \sim}$200 such objects are in the $10$ AU cube at any point in time . Adopting the Pan-STARRS' search volume estimates of \citet{Do2018}, these statistics can be connected to an expected rate for observers on Earth with given detection capability. An average crossing time $t_{cross}\sim 1$ year implies an occurrence rate of objects with $m\le24$ of  2 per year. The two histograms in Figure \ref{fig:seasonalvariation} show the number of ISOs that reach $m\le24$ and $m\le19$  per month, under the constraint that the sun, earth, ISO angle be less than 120$^\circ$. These thesholds are roughly the limiting magnitudes of LSST and Pan-STARRS, although it is important to note that since these telescopes are not capable of seeing the entire night sky, these numbers should be roughly halved to convert to detections. From the plot, LSST will have very little seasonal variability in detections, while Pan-STARRS is more likely to detect these objects in the Spring. This makes sense physically, since at this time, the Earth is moving in the direction of the solar apex, and is thus encountering more of these objects.

\section{On the Determination of Optimal Interception Trajectories}

The kinetic energy imparted via a high-impact ($v\gtrsim40$~km/sec) collision with an impinging ISO would excavate a substantial plume of sub-surface debris. The content of this material could be examined spectroscopically by a companion flyby spacecraft, permitting an ISO's true composition (and the associated consequences for the planet formation process) to be better assessed. This general strategy, which was adopted by the Deep Impact mission \citep{Ahearn2005} requires far less propulsive energy than velocity-matching or sample-return missions, and requires only that the ISO trajectory is determined in sufficient time for an interceptor to be marshaled.

To simplify our analysis, we assume that launches onto interception trajectories are mounted with a single impulsive velocity change, $\Delta V$, from a phase space point belonging to a circular, heliocentric, $a=1$~AU orbit. Energetically, this task is very similar to the launch of a body at rest (in the co-rotating frame) from Earth's $L1$ Lagrange Point \citep{MurrayDermott1999}. In the calculations below, we assume that Earth's gravity is negligible, and that the orbital trajectories of both the ISO and the spacecraft, are Keplerian conic sections. This approach amounts to assuming that early stages of the rocket are used to lift the mission stage up out of Earth's gravitational potential well.  We have  numerically checked that the trajectories of our resulting optimal orbits are not sensitively dependant on the initial conditions, by verifying that the same impulsive thrust sent from L1 and the Earth, under the influence of of Earth's gravitational field, result in endpoints that are closer than $.1\,{\rm AU}$ apart.

With these constraints, we can formulate the problem in the following manner. Given two initial position and velocity vectors at time $t_0$, $\bf{r_0}'$, $\bf{r_0}$, $\bf{v_0}'$, and $\bf{v_0}$, for an interceptor and target, respectively, we ask for the optimal flight time $\Delta t$ and the impulsive change to the initial velocity, ${\bf\Delta v_0}'$, that simultaneously minimize the energy  $E=\frac12 |\Delta \bf{v_0}'|^2$, while ensuring that a collision occurs.

For the specific case of two elliptical orbits, \citet{Leeghim2013} developed a robust optimization algorithm to the optimal trajectory problem using Lagrange multipliers. Here we generalize this approach for elliptical or hyperbolic trajectories for both target and interceptor.

Following \citep{Chobotov1991,Bate1971}, it is useful to adopt the so-called universal variable, defined as

\begin{equation}\label{eq:chi}
\chi =
\begin{dcases}
    \sqrt{a}E & e <1 \\
    \sqrt{-a}F & e>1 \\
\end{dcases} 
\, ,
\end{equation}
given an orbit's semi-major axis, $a$, and eccentric anomaly, $E$, or hyperbolic eccentric anomaly, $F$. In this formalism, the semi-major axis for the hyperbola has a negative value.

Given a value of $\chi$ at a time $t_0+\Delta t$, Kepler's equation can be written,
\begin{equation}\label{eq:kepler}
    \sqrt{\mu}\Delta t = \frac{{\bf r_0\cdot v_0}}{\sqrt{\mu}}\chi^2C(\alpha \chi^2)+(1-\alpha r_0)\chi^3S(\alpha \chi^2)+r_0\chi\, ,
\end{equation}
where $\alpha = 1/a$, $\mu=GM$, and $C(z)$ and $S(z)$ are Stumpff functions, defined as

\begin{equation}\label{eq:stumpff1}
    S(z) = 
    \begin{dcases}
    \frac{\sqrt{z}-\sin\sqrt{z}}{\sqrt{z}^3} & z >0 \\
    \frac{\sinh\sqrt{-z}-\sqrt{-z}}{\sqrt{-z}^3} & z <0 \\
    \end{dcases}     
    \, ,
\end{equation}
and
\begin{equation}\label{eq:stumpff2}
    C(z) = 
    \begin{dcases}
    \frac{1-\cos\sqrt{z}}{\sqrt{z}} & z >0 \\
    \frac{\cosh\sqrt{-z}-1}{-z} & z <0 \\
    \end{dcases}     
    \, .
\end{equation}
In this formalism, the four Lagrange coefficients,

\begin{equation}\label{eq:f}
    f = 1-\frac{\chi^2}{r_0}C(\alpha\chi^2)\, ,
\end{equation}

\begin{equation}\label{eq:g}
    g = \Delta t -\frac{1}{\sqrt{\mu}}\chi^3S(\alpha\chi^2)\, ,
\end{equation}
\begin{equation}\label{eq:fdot}
    \dot{f}=\frac{\sqrt{\mu}}{rr_0}[\alpha \chi^3 S(\alpha\chi^2)-\chi]\, ,
\end{equation}
and
\begin{equation}\label{eq:gdot}
    \dot{g}=1-\frac{\chi^2}{r}C(\alpha\chi^2)\, ,
\end{equation}
uniquely define the position and velocity vector of an orbit after a time $\Delta t$ via,

\begin{equation}\label{eq:position}
    {\bf r}=f {\bf r_0}+g{\bf v_0} \, ,
\end{equation}
and
\begin{equation}\label{eq:velocity}
    {\bf v}=\dot{f}{\bf r_0}+\dot{g}{\bf v_0}\, .
\end{equation}
We define two functions, $\eta_1$ and $\eta_2$, that express the orbital motion of the target and interceptor in terms of their respective universal variables,

\begin{equation}\label{eq:eta1}
\begin{split}
\eta_1(\chi,\Delta t) = \frac{{\bf r_0}\cdot {\bf v_0}}{\sqrt{\mu}}\chi^2C(\alpha \chi^2)\\+(1-\alpha r_0)\chi^3S(\alpha \chi^2)+r_0\chi-\sqrt{\mu}\Delta t \, ,
\end{split}
\end{equation}
and
\begin{equation}\label{eq:eta2}
\begin{split}
\eta_2(\chi',{\bf \Delta v_0'}, \Delta t) = \frac{{\bf r_0'\cdot (v_0'+\Delta v_0')}}{\sqrt{\mu}}\chi'^2C(\alpha \chi'^2)\\+(1-\alpha r_0')\chi'^3S(\alpha \chi'^2)+r_0'\chi'-\sqrt{\mu}\Delta t \, .
\end{split}
\end{equation}
The problem amounts to finding values, $\bf{\Delta v_0'}$, and $\Delta t$ that satisfy,
 
 \begin{equation}\label{eq:equalpos}
     {\bf r'} - {\bf r}=0\, ,
 \end{equation}
 \begin{equation}\label{eq:eta10}
    \eta_1(\chi,\Delta t) = 0\, ,
 \end{equation}
 and
 \begin{equation}\label{eq:eta20}
     \eta_2(\chi',{\bf \Delta v_0'}, \Delta t)=0\, ,
 \end{equation}
 subject to the constraint that 
 \begin{equation}\label{eq:constraint}
     J = \frac12 |{\bf \Delta v_0'}|^2\, ,
 \end{equation}
 is minimized. 
 
 Numerical solutions for this system of equations are straightforward to find by the following iterative process:
 \begin{enumerate}
     \item Choose a trial flight time $\Delta t$.
     \item Solve for roots of the transcendental Equation \ref{eq:eta1} for the universal variable of the target $\chi$ corresponding to $\Delta t$.
     \item Solve the system of four transcendental equations defined by Equation \ref{eq:eta2} and Equation \ref{eq:equalpos} for all three components of ${\bf \Delta v_0'}$ and $\chi'$.
     \item Repeat process until a $\Delta t$ that minimizes Equation \ref{eq:constraint} is found.
 \end{enumerate}

For the purpose of this paper, where a reasonable range of possible flight durations are known \textit{a priori}, the foregoing iterative procedure is sufficient. A more general algorithm is, however, useful, so we present one in the appendix.
\begin{figure}
\begin{center}
\resizebox{0.49\textwidth}{!}{\includegraphics*[trim={.30cm .05cm .85cm .8cm},clip]{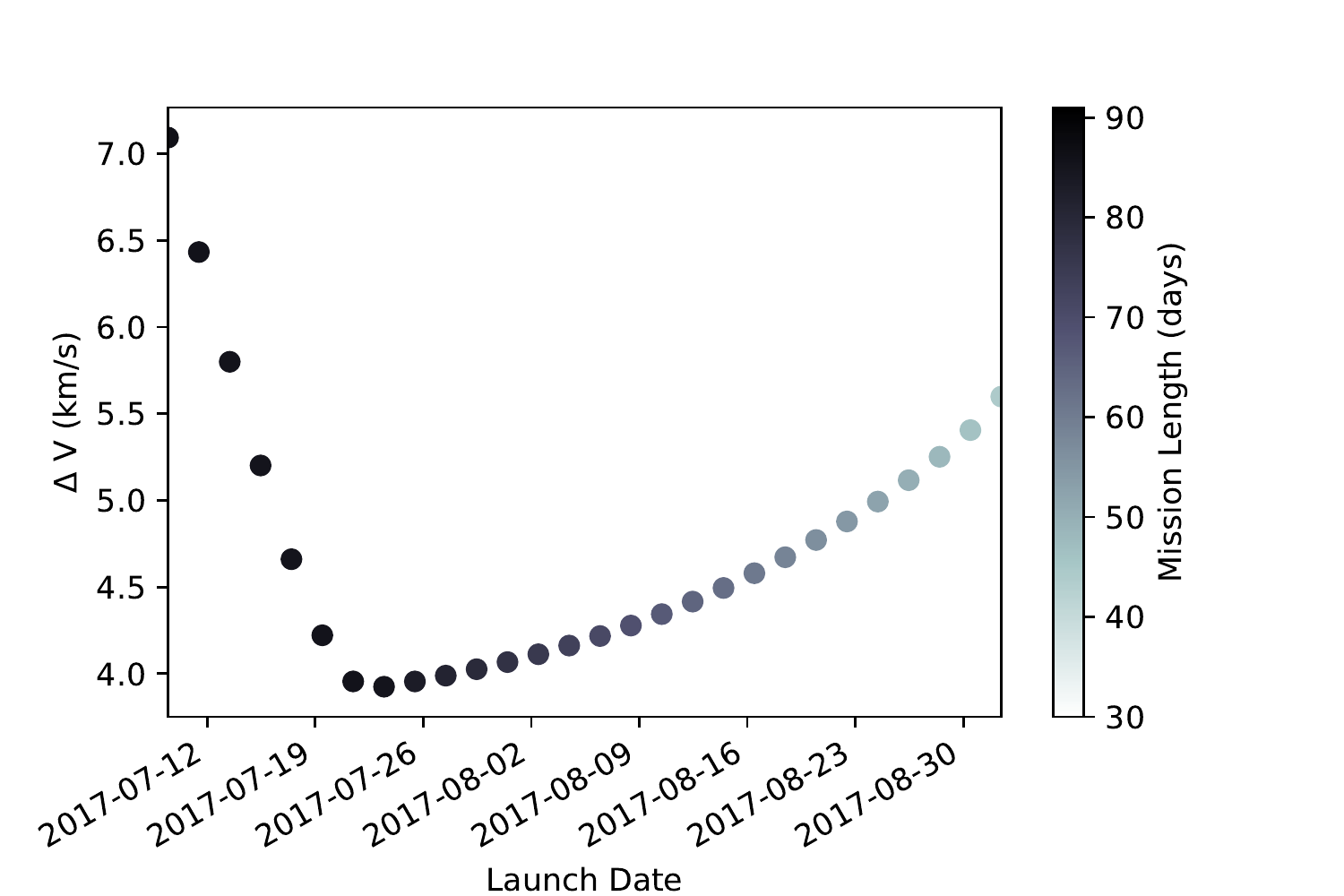}}
\caption{Optimal flight $\Delta V$'s as a function of launch date during the mid-2017 launch window for `Oumuamua interception trajectories. The $x$-axis shows the launch date and the $y$-axis shows the minimum required impulsive change in velocity (relative to Earth's orbit, with Earth treated as a test particle) to intercept `Oumuamua. Each launch opportunity is shaded by the time of flight required for the minimum-energy trajectory. The color-bar on the right is shown in units of days. As a rule of thumb, the minimum energy trajectories lead to impact with the target as it reaches its closest approach to the Earth. }
\label{fig:deltav}
\end{center}
\end{figure}

\begin{figure*}
\begin{center}
\resizebox{0.99\textwidth}{!}{\includegraphics*[trim={.30cm .15cm .45cm .8cm},clip]{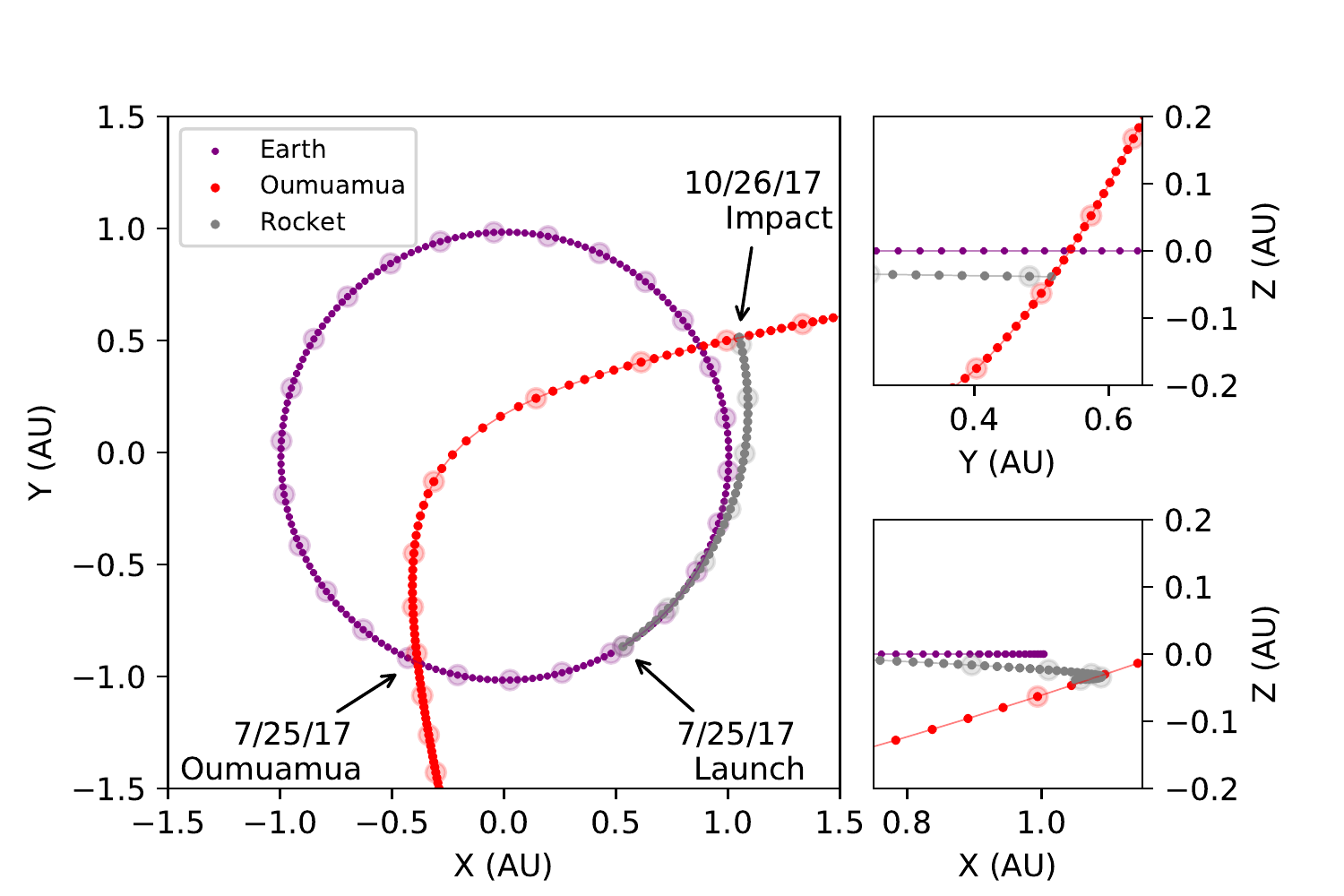}}
\caption{Trajectory of the minimum-$\Delta V$ mission interception mission sent on July 25th 2017, which had a flight time of 83.38 days. The trajectories for `Oumuamua, the Earth, and the rocket are plotted in red, blue and grey respectively in four day intervals in the smaller circles, while the larger circles are plotted in 28 day intervals. The arrows indicate the positions in space of `Oumuamua and the rocket on the launch and interception date, 7/25/2017 and 10/16/2016. Projections in the X-Y,  Y-Z and X-Z planes are shown in the left, right upper, and right lower panel respectively.}
\label{fig:Orbit}
\end{center}
\end{figure*}
\subsection{A Specific Example: `Oumuamua}

`Oumuamua was discovered on 19 Oct. 2017, well after experiencing its periastron passage on 9 Sep. 2017, and indeed, when its heliocentric distance exceeded $r=1$\,AU. Given the late discovery date, and the subsequent span of inaction, it would be extremely difficult to reach. It is interesting, however, to use the above-described procedures to evaluate the mission trajectories that \textit{could} have been employed over a range of launch dates assuming a sufficiently early detection. 

Figure \ref{fig:deltav} was produced using the procedure described above, and charts the optimal values of $\Delta V(t)$ (from Earth's orbit, treating Earth as a test particle) as function of launch dates ranging from 7 July, 2017 to 30 August 2017. Each point on the chart is colored by the time of flight for the optimal trajectory.

It is clear that, given sufficient warning, it would not have been energetically difficult to intercept `Oumuamua. Missions launched during the first several weeks of July 2017 require $\Delta v \approx 4 \,{\rm km/s}$. Figure \ref{fig:Orbit} shows the trajectory of the Earth, the interceptor, and `Oumuamua for the July 25th, 2017 minimum-$\Delta V$ mission  (which had a flight time of 83.38 days). We have verified these matched-conic trajectories with full integrations of the Solar System using the Rebound code \citep{Rein2012}, and assuming departure trajectories from the Earth-Sun L1 point. Moreover, it is evident that as a first-order rule of thumb, the minimum energy trajectory will have the date of impact be when the target reaches closest approach to the Earth. 

\subsection{ Missions to Future Interstellar Asteroids}

\begin{figure}
\begin{center}
\resizebox{0.49\textwidth}{!}{\includegraphics*[trim={.30cm .15cm .0085cm .008cm},clip]{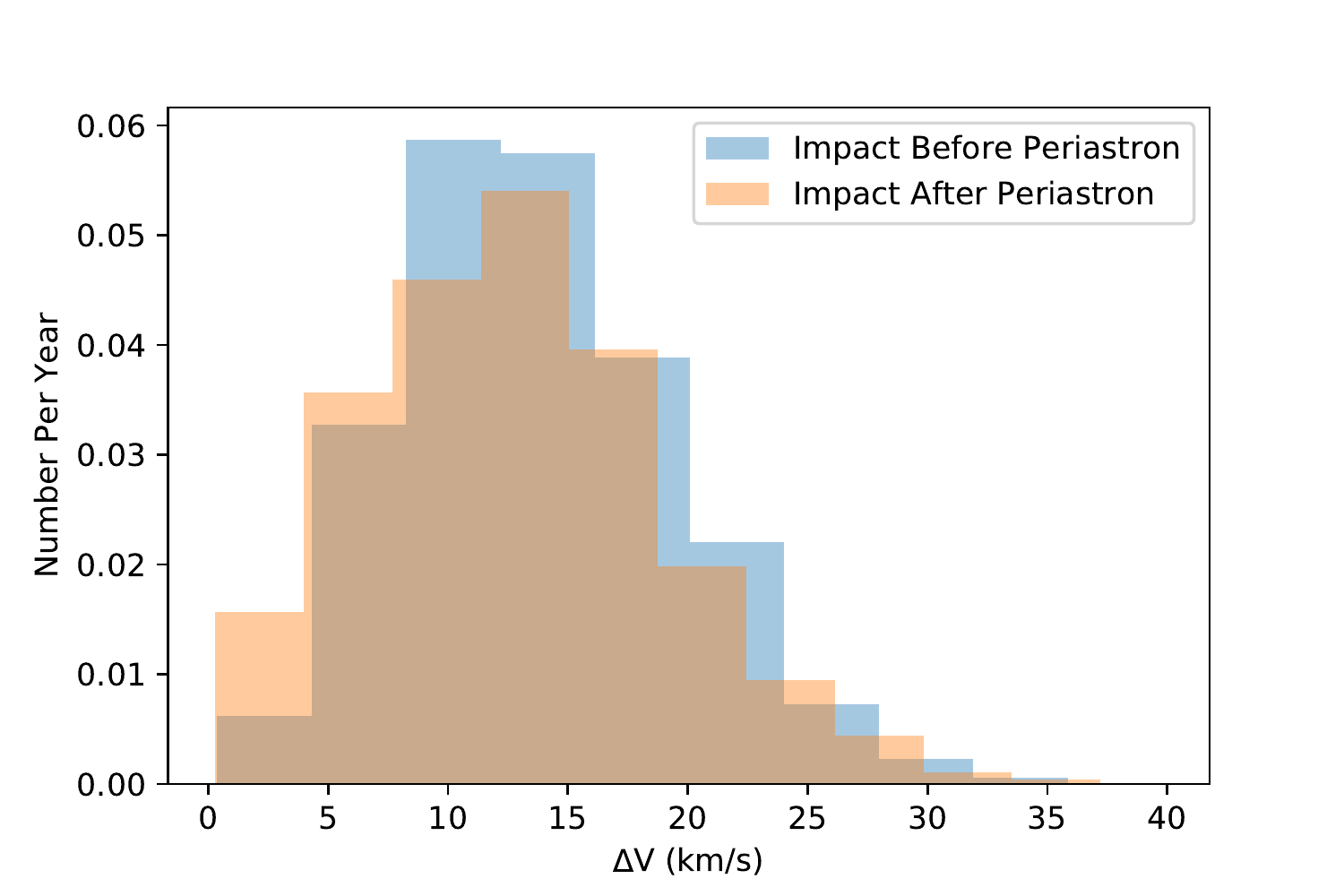}}
\caption{ The required $\Delta V$ and yearly occurrence rates for favorable impactor missions to future ISOs. We only consider trajectories from the Monte Carlo integration in \S 3 that  (i) are visible using LSST with both magnitude and angle constraints, (ii) have a perihelion distance that is closer than 1AU, and (iii) are visible prior to the proposed impact dates, a population that is $\sim1/8$ times the number that are visible by LSST, as shown in Figure \ref{fig:seasonalvariation}. The blue/yellow histograms correspond to missions that impact when the ISO enters/exits the 1AU sphere centered at the Sun. The median $\Delta V$ required for the interception missions is $\sim 12$ km/s. We predict wait times of order $10$ years between favorable mission opportunities with the detection capablities of LSST.}
\label{fig:hypotheticalmissions}
\end{center}
\end{figure}

We now turn to an evaluation of prospects for feasible interception missions to future ISOs, drawing on the integrations that comprise the Monte Carlo simulation described in \S 3. The specific example of `Oumuamua suggests that optimal interception missions reach impact close to the moment when the ISO either enters or exits the 1 AU sphere centered at the Sun. The $\Delta$V required for such an interception mission can roughly be estimated as twice the $\Delta$V required to traverse the distance from the Earth to the asteroid at these dates, given a warning time, defined as the difference in the date of detection and the date of impact. Therefore, we only consider ISO trajectories in the Monte Carlo simulation that (i) are visible using LSST (given both magnitude and angle constraints), (ii) have a perihelion distance that is closer than 1 AU, and (iii) are visible prior to the proposed impact dates. This population is $\sim1/8$ the total number that are visible by LSST, as shown in Figure \ref{fig:seasonalvariation}. Figure \ref{fig:hypotheticalmissions} shows histograms of the required $\Delta V$ for impactor missions prior to and after periastron for each ISO trajectory in this population. The $y$-axis is weighted by the yearly occurrence rates for each $\Delta V$ bin. We find that for the population that satisfies constraints (i)-(iii), the date of detection is always prior to the date that the ISO enters the 1AU sphere, and that the median $\Delta V$ required for the interception missions is $\sim 12$ km/s.

An order-of-magnitude estimate follows quickly from this estimate of the expected distibutions of $\Delta V$ for such a mission. The SpaceX Falcon Heavy quotes a payload capability to Mars of 16,800 kg, which we  conservatively use for the payload constraint to L1. The Deep Impact mission to Tempel I had an impactor weighing $\sim 400$ kg and a scientific package weighing  $\sim 600$ kg \citep{Ahearn2005}. Due to the uncertainty of the position of the ISO, it seems appropriate to use $\sim 16$ impactors, with a total weight of $400$ kg. The mission program is greatly assisted by the expected ~40 km/s velocity of impact with the hyperbolic ISO. Assuming that the remainder of the payload consists of fuel and oxidants, to account for the oxidants and efficiency of the rocket, we allow $\sim 1200$ kg of fuel (with specific energy similar to compressed hydrogen) to produce the $\Delta V$. Equating the kinetic energy to the energy produced by the fuel, we calculate that a maximum $\Delta V \sim 15$ km/s should be attainable, to impart the same amount of kinetic energy (per impact) as the Deep Impact Tempel I interception did. This capability is above the median $\sim 12$ km/s of impact missions before periastron in Figure \ref{fig:hypotheticalmissions}. Given the population of favorable ISO trajectories and order of magnitude assessment of $\Delta V$ capabilities, we predict wait times of order $10$ years between favorable mission opportunities with the detection capablities of LSST. 

\section{Discussion}

`Oumuamua departed as quickly as it arrived, leaving both insights and unanswered questions in its wake. 

The developing opinion prior to `Oumuamua's detection \citep{Jura2011, Moro2009} was that incoming planetesimals with genuinely hyperbolic orbital elements would be a rare occurrence, and upper limits on the number densities from surveys including the Catalina Sky Survey \citep{Drake2009} and the Pan-STARRS1 projects \citep{Chambers2016} have become progressively more restrictive \citep{Engelhardt2014}. Moreover, it was expected that when such objects did appear, they would behave like comets. Our analysis suggests that for `Oumuamua, the combination of a rapid approach, a conductive, extremely cold interior, chaotic tumbling, and an elongated figure provided the subsurface ice with protection from the briefly intense surface heating that transpired during its hyperbolic encounter with the Sun.

It is not yet clear if `Oumuamua's size and shape are particularly indicative of the usual properties of bodies from its source population, and it seems likely that a resource such as Pan-STARRS will detect kindred objects only every few years. LSST, however, with its 2019 first light,  is likely to up the discovery rate to several per year, based on the work presented here and in \citet{Trilling2017}. Additional ISOs, if and when they are found, will set `Oumuamua into a much more definitive context.

Over the past several decades, targeted missions to comets, asteroids, and outer Solar System bodies have provided a raft of important insights into the sequence of events that molded the Solar System. By extension,
there is significant scientific motivation for investigating interstellar asteroids in close proximity. Given the large velocities involved, both orbit-matching and sample-return missions to a passing ISO are likely to be extremely challenging. Our analysis indicates, however, that when LSST is on line, the wait time between opportunities to intercept an ISO with a kinetic impactor is of order  ten years or less.

\section{Acknowledgements}
 This material is based upon work supported by the National Aeronautics and Space Administration through the NASA Astrobiology Institute under Cooperative Agreement Notice NNH13ZDA017C issued through the Science Mission Directorate. We acknowledge support from the NASA Astrobiology Institute through a cooperative agreement between NASA Ames Research Center and Yale University. D. S. thanks the Gruber Foundation and Patricia Gruber for the Gruber Fellowship, and the Stephen B. Butler Scholarship Fund for the Stephen B. Butler Fellowship, both of which also supported the work reported here. We thank Darin Ragozzine, Scott Sandford,  Edward Wright, and J.J. Zanazzi for particularly useful conversations, and Joel Ong for insightful comments.  We also thank the anonymous referee, whose construtive report led to modifications that significantly improved the scientific content of this paper.
\acknowledgements

\appendix
\section{Appendix A}
 Using Lagrange multipliers, \citep{Leeghim2013} demonstrated that for the case of two elliptical orbits, the minimal energy trajectory satisfies,
 
 \begin{equation}\label{eq:lagrange}
     h={\bf \Delta v_0' }^T L_0[(\frac{\partial {\bf r}}{\partial \chi}
     -\frac{\partial {\bf r'}}{\partial \chi'})+\frac{r}{\sqrt{\mu}}
     ({\bf v_0}- {\bf v_0'} - {\bf \Delta v_0'})]=0\, ,
 \end{equation}
 where 
 \begin{equation}\label{eq:Lmat}
     L_0=(\frac{\partial {\bf r'}}{\partial \Delta {\bf v_0'}} -\frac{1}{r'}\frac{\partial {\bf r'}}{\partial \chi'}\frac{\partial \eta_2}{\partial {\bf \Delta v_0'}})^{-1}\, ,
 \end{equation}
 is a 3x3 matrix, and all partial derivatives for the elliptical case are given in Appendix A of \citet{Leeghim2013} \footnote{Note that we have elected to use primed and unprimed variables to denote the target and interceptor, while \citet{Leeghim2013} use overbars.}. Equations \ref{eq:lagrange} and \ref{eq:Lmat}  hold in the general case of hyperbolic or elliptical orbits, but the partial derivatives are different if either or both  orbits are hyperbolic. The hyperbolic Lagrange coefficients are 
 \begin{equation}\label{eq:fhyp}
    f = 1 - \frac{(-a)}{r_0}[\cosh{\Big(\frac{\chi}{\sqrt{-a}}}\Big)-1]\, ,
\end{equation}
and
\begin{equation}\label{eq:ghyp}
    g = \Delta t - \frac{(-a)}{\sqrt{\mu}}[\sqrt{-a}\sinh{\Big(\frac{\chi}{\sqrt{-a}}\Big)}-\chi]\, .
\end{equation}
 
 In order to solve for $\frac{\partial {\bf r}}{\partial \chi}$ and $\frac{\partial {\bf r'}}{\partial \chi'}$, the partial derivatives in the hyperbolic case are

\begin{equation}\label{eq:partialf}
    \frac{\partial f}{\partial \chi}=-\frac{\sqrt{-a}}{r_0}\sinh{\Big(\frac{\chi}{\sqrt{-a}}\Big)}\, ,
\end{equation}
and
\begin{equation}\label{eq:partialg}
    \frac{\partial g}{\partial \chi}=-\frac{(-a)}{\sqrt{\mu}}(\cosh{\Big(\frac{\chi}{\sqrt{-a}}\Big)}-1)\, .
\end{equation}
 
 To solve for the matrix $\frac{\partial {\bf r'}}{\partial \Delta {\bf v_0'}}$,
 \begin{equation}
    \frac{\partial f}{\partial a} = -\frac{1}{r_0'}[1-\cosh{\Big(\frac{\chi'}{\sqrt{-a}}\Big)}
    +\frac{\chi'}{2\sqrt{-a}}\sinh{\Big(\frac{\chi'}{\sqrt{-a}}\Big)}]\, ,
\end{equation}
and
\begin{equation}
    \frac{\partial g}{\partial a}=\frac{-1}{\sqrt{\mu}}[\chi' -\frac{3\sqrt{-a}}{2}\sinh{\Big(\frac{\chi'}{\sqrt{-a}}\Big)}+\frac{\chi'}{2}\cosh{\Big(\frac{\chi'}{\sqrt{-a}}\Big)}]\, .
\end{equation}
 
 Finally, the vector $\frac{\partial \eta_2^h}{\partial \Delta v_0'}$ for the hyperbolic case is,
 \begin{equation}
    \frac{\partial \eta_2^h}{\partial \Delta v_0'}=\frac{\partial \eta_2^h}{\partial a}
    \frac{\partial a}{\partial \Delta v_0'}+\frac{-a}{\sqrt{\mu}}[\cosh{\big(\frac{\chi'}{\sqrt{-a}}\big)}-1]{\bf r_0'}^T\, ,
\end{equation}
where
\begin{equation}
    \frac{\partial a}{\partial \Delta v_0'}=\frac{2a^2}{\mu}({\bf v_0'+\Delta v_0'})^T \, ,
\end{equation}
and  the scalar is 

\begin{equation}
\begin{split}
    \frac{\partial \eta_2^h}{\partial a}=\chi' - \frac32 \sqrt{-a}\sinh{\big(\frac{\chi'}{\sqrt{-a}}\big)}
    +\frac{\chi'}{2}\cosh{\big(\frac{\chi'}{\sqrt{-a}}\big)}
    +\frac{{\bf r_0'} \cdot ({\bf v_0'+\Delta v_0'})}{\sqrt{\mu}}[1-\cosh{\big(\frac{\chi'}{\sqrt{-a}}\big)}+\frac{\chi'}{2\sqrt{-a}}\sinh{\big(\frac{\chi'}{\sqrt{-a}}\big)}]\\
    +\frac{r_0}{2}[\frac{-1}{\sqrt{-a}}\sinh{\big(\frac{\chi'}{\sqrt{-a}}\big)} +\frac{\chi'}{-a}\cosh{\big(\frac{\chi'}{\sqrt{-a}}\big)}]
\end{split}\, .
\end{equation}

Given these definitions, a robust algorithm for finding the optimal trajectory for elliptical or hyperbolic orbtis is as follows:

 \begin{enumerate}
     \item Choose a trial flight time $\Delta t$.
     \item Solve for roots of the transcendental Equation \ref{eq:eta1} for the universal variable of the target $\chi$ corresponding to $\Delta t$.
     \item Solve the system of four transcendental equations defined by Equation \ref{eq:eta2} and Equation \ref{eq:equalpos} for all three components of ${\bf \Delta v_0'}$ and $\chi'$.
     \item Evaluate Equation \ref{eq:lagrange}.
     \item Repeat process until Equation \ref{eq:lagrange} is satisfied. 
 \end{enumerate}

\end{document}